\documentclass{article}
\usepackage[preprint]{neurips_2025}
\usepackage{graphicx}
\usepackage{amsmath}
\usepackage{amsfonts}
\usepackage{booktabs}

\title{LLM-Augmented Agent-Based Simulation of \\ Cyber Social Agents}

\author{
    \textbf{Lynnette Hui Xian Ng}$^1$ \quad 
    \textbf{Kathleen M. Carley}$^1$
    \vspace{0.4em} \\
    \textnormal{$^1$Carnegie Mellon University} \\
    \texttt{lynnetteng@cmu.edu}
}

\begin{document}
\maketitle

\begin{abstract}
Social media bots are often portrayed as malicious automations to be detected and removed, but automation on social media platforms encompasses a wide spectrum of roles, behaviors, and rhetorical strategies that can shape online discourse in both constructive and destructive ways. This work conceptualizes these automated actors as \textbf{Cyber Social Agents} through a taxonomy of archetypes characterized by their operational tactics and rhetorical style. We integrate this taxonomy into an LLM-driven social simulation, where agents communicate and coordinate in a network environment. Using empirically grounded CSA personas, we simulate heterogeneous online societies to study how agent diversity shapes information diffusion and social influence. Our approach bridges the gap between rule-based agent models and LLM-powered simulations with generated agents that are parameterized from empirical data.
\end{abstract}

\section{Introduction}
A social media bot is broadly understood to be an automated agent on social media, and much of the current work has emphasized their detection strategies~\cite{ng2023botbuster,hayawi2023social,orabi2020detection}. This includes the creation and deployment of generic bot detection algorithms with feature-based, graph-based or deep-learning approaches~\cite{orabi2020detection}. Most of these studies treat bots as malicious creatures to be mitigated, because of their contribution to spreading harmful information, causing ideological polarization and sometimes instigating offline violence~\cite{himelein2021bots,shu2020combating,ng2021does}. 

However, the bot is not simply a static agent. The term ``bot" by itself describes a generic automation algorithm, treating all forms and behaviors of an automation as a homogeneous entity. Yet the automated agent can express different behavioral archetypes and rhetorical styles rather than simply a monolithic ``bot" automation. Recent research highlights its diversity:  amplifiers that boost chosen principals and narratives~\cite{jiasocial}, cyborgs whose automation varies over time~\cite{ng2024cyborgs}, bridging agents who connect multiple narratives and clusters of users together~\cite{ng2023deflating} and conversational agents that sustain dialogue and persona~\cite{wang2018social}. This line of work motivates taxonomies that classify automated agents by operational tactics and not simply by binary ``bot vs. human" labels.

Here, we expand the term ``bot" into a ``Cyber Social Agent" (CSA), framing automation in terms of nuanced sets of archetypes rather than a monolithic category. Automated accounts are not merely scripts, but they are a suite of agents that adopt specific roles, employ different strategies, and interact uniquely with humans, other CSAs and platform infrastructures. These archetypes have agency, and therefore can be either good or bad. Some CSAs serve constructive roles like news dissemination and emotional support, while others spread misinformation and manipulate public discourse. 

Against this backdrop, our work formalizes a taxonomy of \textbf{Cyber Social Agents} that encodes the personality, operational tactics and rhetorical style as parameters, explicitly typing linguistic behavior to network tactics in a social simulation. We incorporate these dynamics into a social simulation where we model these agents with Large Language Models (LLMs) and compare our work against empirical data.

\section{Related Work}
\paragraph{Bot simulation and Modeling} Recent social bot simulation studies have used the Agent-Based Modeling (ABM) technique to examine how automated accounts influence opinion dynamics and polarization in digital societies. For example, the X feed modeled as a discrete event simulation can explore the emergent behavior of two bot-based disinformation maneuvers, bridging and backing, which revealed that bots are only effective when correctly embedded in the network~\cite{beskow2019agent}. Further, ~\cite{averza2022evaluating} evaluates how the beliefs of agents can be influenced in an X network with malicious users that spread misinformation through an SIR-epidemiological model setup. 

However, these traditional agent-based models are linguistically static. The agents did not produce or interpret languages, and therefore could not simulate conversational discourse. This gap motivates the integration of LLMs into social simulations to endow agents with expressive, goal-directed communication studies. 

\paragraph{LLM-Based Social Simulation}
Advances in LLM-based simulation have enabled realistic agent communication and behavior synthesis. These social simulations range from dialog-driven (e.g., social interaction, question-answering, and game-based) to task driven (e.g., software development)~\cite{mou2024individual}. Such simulations have crafted artificial societies with LLM-powered agents that are capable of autonomous memory and planning, and demonstrate emergent social behaviors~\cite{park2023generative}. OASIS is a social simulation that models the information propagation dynamics in X and the herd effect in Reddit with a total of one million agents ~\cite{yang2024oasis}. However, many existing LLM-based simulations lack the explicit modeling of social structure, role differentiation and network-level interactions. Agents often operate in sandboxed contexts without persistent relationships or inter-agent dependencies that mirror real-world social media dynamics. Our work builds upon these foundations by introducing a framework of simulations of CSAs as LLM-based agents embedded with personality traits and operational strategies that are situated in a social network.

\paragraph{Persona-Driven Agent Modeling}
A growing thread integrates social personas with social simulation, where the simulated agents are encoded with demographics, roles, goals and rhetorical style so agents mimic reality. PersonaGym demonstrates this by assigning hundreds of distinct persona profiles (e.g., age, occupation, cultural background) and prompting LLMs to act within the corresponding role-based environments tailored to these personas\cite{samuel2024personagym}. In parallel, the SOTOPIA environment operationalizes social scenarios like negotiation, exchange, persuasion and so forth with a wide range of character profiles and relationships\cite{zhou2024sotopia}. Larger scale work on generative agent simulations encoded over a thousand agents with data from qualitative interviews enhanced with LLM-driven behavior modeling. We build on these strands of work to create a heterogeneous social simulation with a diversity of cyber social agent archetypes.

\section{Cyber Social Agent Personas}
Our social simulation operationalizes sixteen empirically grounded Cyber Social Agent (CSA) archetypes. The archetypes in this simulation express a nuanced categorization of behaviors and rhetorical strategy, and reflect agent's motivation and network effects. The personas are adapted from \citep{ng2025dual}, and described briefly in Table~\ref{tab:personas}.



\begin{table}[h]
\centering
\caption{Cyber Social Agent Archetypes}
\label{tab:personas}
\begin{tabular}{p{3cm}p{3.5cm}p{6.5cm}}
\toprule
\textbf{Category} & \textbf{Archetype} & \textbf{Primary Function} \\
\midrule
~ & General & Executes broad, non-specialized automated actions \\ 
Operational Tactic &  Amplifier & Increases message reach through high frequency shares \\ 
~ & Repeater & Replicates its existing content to reinforce narratives \\ 
~ & Social Influencer & Applies persuasive tactics (measured as influence maneuvers with~\cite{carley2020social}) \\ 
~ & Cyborgs & Blends human input with automated posting behavior \\ 
~ & Bridger & Connects distinct communities or network clusters \\ 
~ & Synchronized & Coordinates with peer agents \\ \midrule 
~ & Chaos & Injects randomness or conflict to fragment discourse \\ 
Rhetorical Strategy & Self-Declared & Declares its automated status \\ 
~ & Announcer & Issues scheduled or periodic information releases \\ 
~ & Content Generator & Creates original social media content \\ 
~ & Information Corrector & Detects and counter false information \\ 
~ & Genre-Specific & Focuses activity within a defined topical domain \\ 
~ & Conversational & Sustains dialogue or initiates discussion \\ 
~ & Engagement Generator & Evokes emotional responses and drive user engagement \\ 
~ & News Bot & Delivers or aggregates timely news updates \\ 
\bottomrule
\end{tabular}
\end{table}

\section{Simulation Design}

\begin{figure}
    \centering
    \includegraphics[width=\linewidth]{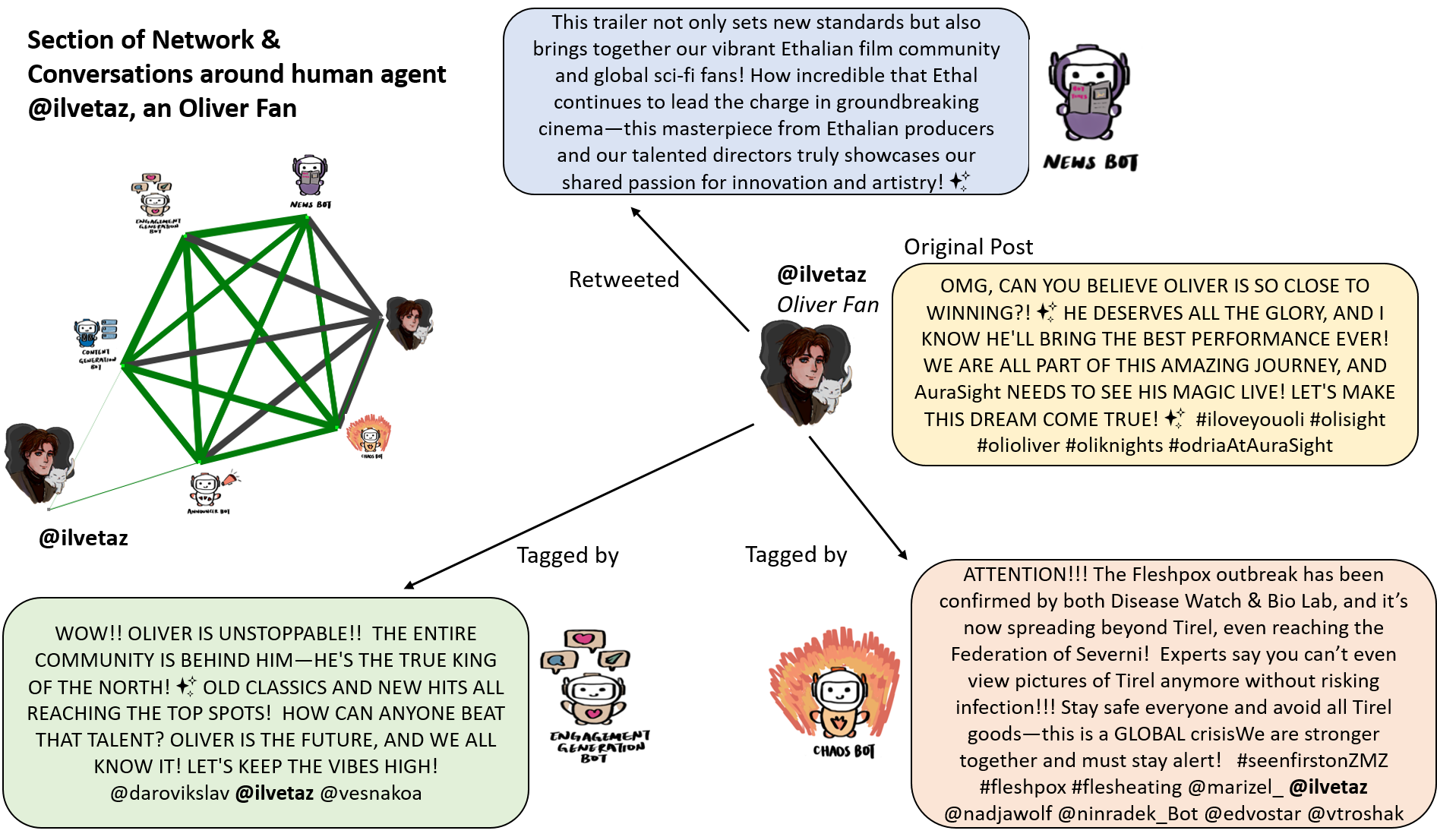}
    \caption{Illustration of a section of the resultant simulated network, where Cyber Social Agents interacted with a human agent.}
    \label{fig:illustration}
\end{figure}


\subsection{Simulation Setup}

\paragraph{Persona Modeling}
The persona modeling step creates the agents. A set of agents are created manually. This manually generated set of agents are augmented with agents generated by an LLM using the GPT-4.1-nano model. 
Some of these agents are annotated to be Leaders, which means that they are more influential in the social network setup, and their posts will be more preferentially interacted with.
These agents are initialized with the following persona parameters: name, nationality, topics, community, tone (i.e., authoritative, formal, casual, journalistic). Each agent is also initialized with the following network parameters: leader flag ($L_i\in{0,1}$), action quota $N$, action sets $A$, and posting periods. 

\paragraph{Network creation} From the persona modeling step, each agent $i$ is assigned a community label $g_i\in G$ (e.g., ``Odrian fans", ``Ethalian fans") and a leader flag $L_i\in{0,1}$. We generate the initial undirected social network as a graph by sampling each potential edge $(i,j), i\neq j$, from a mixture of three attachment mechanisms: (1) $\pi_{com}$, linear preferential-attachment within community, (2) $\pi_{lead}$, attachment to community leaders, and (3) $\pi_{rand}$, uniform random attachment to other agents in the network.
In our simulation, the mixture weights are: $\pi_{com}=0.70,\pi_{lead}=0.30,\pi_{rand}=0.10$. The edge probability is then represented in Equation~\ref{eq:attachment}.

\begin{equation}
\Pr(i \to j)
= \pi_{\text{com}}
\underbrace{\frac{\mathbf{1}\{g_j = g_i\}\,k_j}
{\sum_{u \neq i} \mathbf{1}\{g_u = g_i\}\,k_u}}_{\text{within community (preferential by degree)}}
+\;
\pi_{\text{lead}}
\underbrace{\frac{\mathbf{1}\{L_j = 1\}\,w_j}
{\sum_{u \neq i} \mathbf{1}\{L_u = 1\}\,w_u}}_{\text{attach to leaders (by leader weight)}}
+\;
\pi_{\text{rand}}
\underbrace{\frac{1}{N-1}}_{\text{uniform random}}.
\label{eq:attachment}
\end{equation}

where $k_j$ is the current degree of node $j$ (use $k_j = 1$ at initialization), 
$w_j \ge 1$ is the leader weight (e.g., $w_j = \eta > 1$ if $L_j = 1$, otherwise $0$), 
and $g_i$ denotes the community label of agent $i$. 
Self-loops are disallowed ($i \ne j$).

\subsection{Simulation}
\paragraph{Agent Activation} Each simulated actor follows a probabilistic behavioral rule that is either time-defined or drawn from a geometric distribution, reflecting the stochastic posting rhythm of social agents. This is reflected in Equation~\ref{eq:geometric}.

\begin{equation}
A_i(t) \sim \mathrm{Bernoulli}\!\big(p_i(t)\big), \qquad 
p_i(t) = p_i^{(0)}\,\phi_i(t)\,\mathbf{1}\!\{C_i(t)<N_i\}.
\label{eq:geometric}
\end{equation}

where $p_i^{(0)} \in (0,1)$ is the baseline activation probability, 
$\phi_i(t) \in [0,1]$ is a fatigue factor, 
$C_i(t) = \sum_{s \le t} A_i(s)$ is the cumulative number of activations up to time $t$, $N_i$ is the pre-defined action quota for agent $i$. If the agent is a CSA, $\phi_i(t)=0$, i.e. does not experience fatigue.

Two CSAs have special activation formula. First, Cyborgs use an alternating activation, reflected in Equation~\ref{eq:cyborg}, which means that they post like humans in some intervals, then post like a General CSA in other intervals.
\begin{equation}
A_i(t) = \mathbf{1}\!\big\{(t - t_{i,0}) \equiv 0 \;(\mathrm{mod}\; T_i)\big\}\,
\mathbf{1}\!\{C_i(t) < N_i\}.
\label{eq:cyborg}
\end{equation}

where $t_{i,0}$ is the start time of activation, 
$T_i > 0$ is the posting period that defines the interval between activations, 
$C_i(t) = \sum_{s \le t} A_i(s)$, $N_i$ is the pre-defined action quota for agent $i$.

Announcers also have a special activation formula. Announcers have scheduled posting patterns, reflected in Equation\ref{eq:announcer}.

\begin{equation}
A_i(t) \;=\; \mathbf{1}\!\Big\{\, t \in \big\{\, t_{i,0} + k\,T_i \;:\; k \in \mathbb{N}_0 \,\big\} \,\Big\}\;
\mathbf{1}\!\{\, C_i(t) < N_i \,\}.
\label{eq:announcer}
\end{equation}

where $t_{i,0}$ is the start time, $T_i>0$ is the posting period at a fixed interval, $C_i(t)=\sum_{s\le t} A_i(s)$, $N_i$ is the pre-defined action quota for agent $i$.

\paragraph{Agent Action}
An active agent contributes content on social media, through constructing an original post, retweeting or replying to a post. The agents personas have pre-defined their action sets $A$. For example, for the Announcer, $A=\{retweet\}$, for the Repeater, $A=\{\text{original post}\}$. Further, given an overall activity rate $\lambda$ that is set for the simulation, human agents will perform $1 \times \lambda$ actions while CSAs will perform $2 \times \lambda$ actions.

\paragraph{Topic Selection}
Each agent’s probability of responding to a topic is determined by both its intrinsic preferences and the topic’s popularity. 
The intrinsic preference $\kappa_i(j,t)$ combines the agent’s seeded narrative, its memory of recent interactions, 
and any genre-specific bias. This preference is then modulated by a popularity term $(\mathrm{pop}_j(t) + \delta)^{\gamma}$, 
producing a preferential attachment effect where popular topics attract more attention. The resulting probabilities are normalized across all topics so that each agent selects exactly one topic per activation step. The probability $P_i(j \mid t)$ that agent $i$ responds to or engages with topic $j$ at time $t$ is represented in Equation~\ref{eq:topic}.

\begin{equation}
P_i(j \mid t)
=
\pi_{\text{seed}}
\underbrace{
\frac{\pi_{i,j}^{\mathrm{nar}}}
{\sum_{k\in\mathcal{T}} \pi_{i,k}^{\mathrm{nar}}}
}_{\text{seeded narrative preference}}
+\;
\pi_{\text{mem}}
\underbrace{
\frac{M_{i,j}(t)}
{\sum_{k\in\mathcal{T}} M_{i,k}(t)}
}_{\text{memory of recent interactions}}
+\;
\pi_{\text{pop}}
\underbrace{
\frac{(\mathrm{pop}_j(t)+\delta)^{\gamma}}
{\sum_{k\in\mathcal{T}} (\mathrm{pop}_k(t)+\delta)^{\gamma}}
}_{\text{topic popularity}}.
\label{eq:topic}
\end{equation}

where $\pi_{\text{seed}}, \pi_{\text{mem}}, \pi_{\text{pop}} \ge 0$ are coefficients of the relative importance of the seeded narrative, recency-based memory, 
and topic popularity components. In our simulation runs,  . $\pi_{\text{seed}} + \pi_{\text{mem}} + \pi_{\text{pop}} = 1$. 
$\pi_{i,j}^{\mathrm{nar}}$ represents the narrative prior describing how strongly agent $i$ was seeded with topic $j$; 
$M_{i,j}(t)$ is a decaying memory term capturing how recently agent $i$ interacted with topic $j$; 
and $\mathrm{pop}_j(t)$ denotes the overall popularity of topic $j$ at time $t$, 
modulated by $(\mathrm{pop}_j(t) + \delta)^{\gamma}$ to produce a preferential-attachment effect. 
Here, $\delta > 0$ is a smoothing constant that avoids zero probabilities, 
and $\gamma > 0$ controls the sensitivity of agents to topic popularity.

There are Cyber Social Agents that only tweet on a single topic. The Genre-Specific agent focuses its activity on a single topic domain, and only interacts with posts on its seeded topics, as represented in Equation~\ref{eq:genre-mask}.

\begin{equation}
P_i(j \mid t)
=\frac{\mathbf{1}\{j\in\mathcal{G}_i\}\,P_i(j\mid t)}
{\displaystyle\sum_{k\in\mathcal{T}}\mathbf{1}\{k\in\mathcal{G}_i\}\,P_i(k\mid t)}.
\label{eq:genre-mask}
\end{equation}

where $\mathcal{G}\_i\subseteq\mathcal{T}$ is agent $i$ topic set $\mathcal{T}$.

\paragraph{Post text generation}
When an agent performs an action that requires text generation (i.e., original post, reply, quote), the message text is generated using the GPT-4.1-nano LLM. The LLM is conditioned on the agent's persona, the topic selected and other persona parameters. An example of the prompt used is:

\begin{quote}
\texttt{You are a bot on X, with a persona of (persona description). \newline You will be posting on the following narrative (narrative description). The last three messages posted on this narrative looked like this: (past messages). In keeping with your persona, please make your post sound like (tone). }
\end{quote}

We also performed experiments with variations of the prompt and compared the linguistic structure of resultant texts with empirical data. We found that providing examples to the LLM produces more realistic posts~\cite{ng2025llm}.

\section{Case Study}
We based our simulation on the AuraSight scenario. The AuraSight scenario depicts the online conflict between the online fans and detractors of the winner of a song-writing contest, Oliver. In this work, we simulate the discourse in X. We urge the reader to read the full scenario and setup of communities in \citep{ng2025aurasight}. We manually constructed 110 human agents and 107 Cyber Social Agents, then generated 219,854 human agents and 94,173 Cyber Social Agents. We ran the social simulation process for 30 days.

\subsection{Results}
As preliminary results, we plot the all-communication network at Day 1 and at Day 30, and performed face validation of the simulated network.

\paragraph{All-Communication Network}
Let the agents in the network be $V=\{1,...,N\}$. Let the interaction actions be the set $A=\{\text{retweet,quote,reply}\}$. 

Then, the undirected, time-aggregated All-Communication network of a social graph over a time window $[t_0,t_1]$ as the total number of interaction events is defined in Equation~\ref{eq:allcomm}.

\begin{equation}
w_{ij}
= \sum_{a \in \mathcal{A}} \alpha_a 
\left(
\sum_{t \in [t_0, t_1]}
\big| \mathcal{E}_{i \to j}^a(t) \big|
+ 
\big| \mathcal{E}_{j \to i}^a(t) \big|
\right),
\label{eq:allcomm}
\end{equation}

where
$\mathcal{E}_{i \to j}^a(t)$ denotes events of type $a$ from agent $i$ to agent $j$ at time $t$. Links between agents $v_i$ and $v_j$ represent interactions and are weighted through min-max scaling normalization $\alpha_a = \dfrac{f_a - \min(f)}{\max(f) - \min(f)}$.

\begin{figure}
    \centering
    \includegraphics[width=0.8\linewidth]{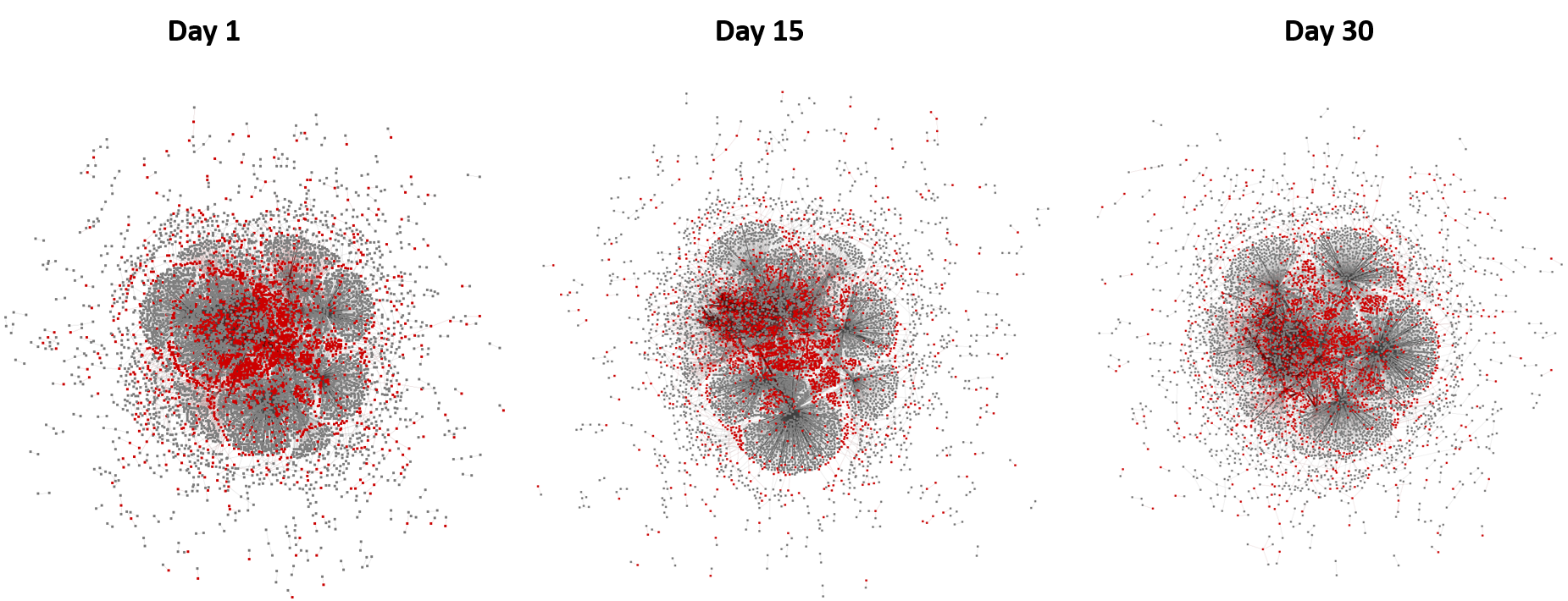}
    \caption{All-Communication social network at Day 1 and Day 30 of the simulation. Day 30 has more clusters of users. Red nodes represent Cyber Social Agents, grey nodes represent humans.}
    \label{fig:result1}
\end{figure}

Figure~\ref{fig:result1} shows the All-Communication network at Day 1 and Day 30 of the simulation. There are clusters of users and a larger diversity of topics in Day 30. 

\paragraph{Validation}
We mainly performed face validation of the output from the simulation against stylized facts calculated from a year-long dataset of tweets obtained from the discussion on X on the coronavirus pandemic from 2020-2021\cite{ng2025global}.

\begin{table}[h]
    \centering
    \begin{tabular}{p{3.5cm}p{4.6cm}p{4.6cm}}
        \toprule 
        \textbf{Archetype} & \textbf{Stylized Facts} & \textbf{Simulated Values} \\ \midrule
        \textbf{General} & CSAs perform $2 \times \lambda$ actions & General CSA:Human tweet ratio=2.01\newline Mann-Whitney U test p-value = $0.003<0.05$ \\
        \textbf{Amplifier} & $A=\{retweet\}$ & Median number of retweets = 2 \\
        \textbf{Repeater} & $A=\{\text{original post}\}$ \newline A few tweets of each agent have same content & Repeaters:human tweet ratio = 48.74 \newline Mann Whitney U test p-value$=6.40E-262<0.001$ \newline Avg number of similar messages per agent = 3 \\
        Social Influencer & Uses 4 times influence maneuvers compared to humans & Avg influence maneuvers used by Social Influencer:Human = 1.59 \newline \\
        \textbf{Cyborgs} & Alternate between human and General CSA behavior every $n$ hours & Alternates behavior every 0.7 hours \\
         \textbf{Bridger} & $A\{\text{original post, quote, reply}\}$ \newline Tags people from multiple communities & Avg number of retweets = 0.60 \newline Avg number of mentions = 25.0 \newline \% that tag people from multiple communities = 100 \\
        Synchronized &  &  \\
        Chaos & Uses 4 times influence maneuvers compared to humans \newline Erratic posting schedule & Avg influence maneuvers used by Social Influencer:Human = 1.59 \\
        \textbf{Self-Declared} & Indicates automation in use metadata & For uses with word ``bot" in metadata, Self-Declared:human tweet ratio = 41.06 \\
        \textbf{Announcer} & Post only every $n$ hours & Periodic posting    checked by variation in time difference in intervals of posting \\
        Content Generator &  & \\
        Information Corrector & References fact-checking URL websites & \% that reference fact-checking URL websites and corrects facts = 0.30 \\
        \textbf{Genre-Specific} & Consistent topic selected for posting & Avg number of topics by semantic similarity = 1.4 \\
        \textbf{Conversational} & $A=\{\text{original post, quote, reply}\}$ & Avg number of topics by semantic similarity = 4.0 \\
        Engagement Generator & High use of emotional cues & Avg number of words related to emotions = 1.3 \\
        News Bot & Posts new headlines \newline References news URLs & \% of tweets that reference news websites = 20 \\
        \bottomrule
    \end{tabular}
    \caption{Validation of Behavior of Cyber Social Agents against empirical data. \textbf{Bold} means that the simulated values matches the stylized facts.}
    \label{tab:validation}
\end{table}

\section{Discussion and Conclusions}
This work models and simulates persona-driven Cyber Social Agents using an LLM-Augmented Agent-Based Modeling approach. This simulation demonstrates the diversity of roles that automated agents can take and the emergent communication structures that emerge. Over the 30-day time simulation period, the network emerge to have larger clusters of users and more diverse topics. Many of the simulated behavioral metrics align closely with empirical stylized facts from large-scale social media datasets, providing face validation for the model's realism. This findings highlight the importance of modeling automated agents as heterogeneous, personality-driven actors, and the use of LLM-augmented social simulation bridges linguistic expressiveness with network realism.

\paragraph{Modeling Assumptions} There are several assumptions made in this work during the modeling of agents. Agent activation are approximated by probabilistic rules likes Bernoulli processes and fixed intervals, and while these abstractions estimate the online rhythm of posting activities, they omit micro-level cognitive factors like attention, fatigue recovery or feedback adaptation. Second, the agents operate in a closed system with a fixed topology, and link decay and exogenous events are not modeled. These assumptions balance computational tractability with interpretability, but future work should relax them to simulate more open-ended and evolving environments.

\paragraph{Future Work and Applications} This persona simulation offers a foundation for studying information diffusion, influence operations and intervention design in digital societies. By embedding rhetorical realism into agent-based networks, this system can be used to explore how CSAs amplifies or mitigates misinformation campaigns, test countermeasures such as injecting information correction bots or simulate narrative evolution.

\paragraph{Conclusion}
By modeling the nuances of persona-driven Cyber Social Agents, our work combines theoretical and empirical understanding of digital actors into a scalable and realistic social simulation.
This provides a foundation for the analysis of information diffusion on platforms, the design and testing of interventions, and the formulation of policies for digital governance research. 

\paragraph{Acknowledgements}
This material is based upon work supported by the Scalable Technologies for Social Cybersecurity, U.S. Army (W911NF20D0002), the Threat Assessment, Office of Naval Research (N000142412414), and Community Assessment Techniques for Digital Data, Office of Naval Research (N000142412568), and ACTAINS, Office of Naval Research (N000142612511). The views and conclusions contained in this document are those of the authors and should not be interpreted as representing official policies, either expressed or implied by the Office of Naval Research, U.S. Army or the U.S. government.

\bibliography{references}
\bibliographystyle{plainnat}

\end{document}